\documentstyle[epsf,rotate]{mn}
\newcommand{\abs}[1]{|#1|}
\newcommand{\be}{\begin{equation}}
\newcommand{\ee}{\end{equation}}
\newcommand{\ptl}[2]{{\frac{\partial #1}{\partial #2}}}
\newcommand{\ptlp}[2]{{\frac{\partial (#1)}{\partial #2}}}
\def\Msol{{M_\odot}}
\def\kpc{{\rm kpc}}
\def\spose#1{\hbox to 0pt{#1\hss}} 
\def\lta{\mathrel{\spose{\lower 3pt\hbox{$\sim$}}\raise 2.0pt\hbox{$<$}}}
\def\gta{\mathrel{\spose{\lower 3pt\hbox{$\sim$}}\raise 2.0pt\hbox{$>$}}}

\begin{document}

\title[Rotating Halos]{Rotating Halos and the Microlensing MACHO Mass Estimate}
\author[Geza Gyuk and Evalyn Gates]{Geza Gyuk$^1$ and Evalyn Gates$^{2,3}$ \\ 
\\
$^1$S.I.S.S.A., via Beirut 2--4, 34014 Trieste, Italy \\
$^2$Adler Planetarium,1300 Lake Shore Drive, Chicago, IL~~60605\\
$^3$Department of Astronomy \& Astrophysics, The University of Chicago, 
Chicago, IL~~60637}
\date{Received ***}
\maketitle

\begin{abstract}
We investigate the implications of a bulk rotational component of the halo
velocity distribution for MACHO mass estimates. We find that for a
rotating halo to yield a MACHO mass estimate significantly below that of
the standard spherical case, its microlensing must be highly concentrated
close to the Sun. We examine two classes of models fitting this criteria:
a highly flattened $1/r^2$ halo, and a spheroid-like population with whose
density falls off as $1/r^{3.5}$. The highly flattened $1/r^2$ models can
decrease the implied average MACHO mass only marginally and the spheroid
models not at all.  Generally, rotational models cannot bring the MACHO
mass implied by the current microlensing data down to the substellar
range.
\end{abstract}


\section{Introduction}
Recent microlensing results from the MACHO collaboration suggest that in
the context of a spherical isothermal model for the Galactic halo, some
significant fraction of the halo is composed of objects with masses of
about $0.4 M_{\odot}$ \cite{MACHOmass}.  Such masses are consistent with
several astrophysical candidates for MACHOs -- low mass main sequence
stars, white dwarfs, and black holes.  Each of these candidates, however,
is subject to a variety of constraints that present serious challenges for
succesful models in which they form a significant fraction of the halo.
Low mass main sequence stars should be easily visible, and recent direct
searches for these stars limit their contribution to the halo to be less
than $3\%$ \cite{halostars,reddwarfs}.  The limits on halo white dwarf
stars are much looser.  Very old white dwarfs would be cool and faint
enough to have evaded direct detection thus far and yet still be present
in the halo in significant numbers
\cite{whitedwarfs,IMFproblem2,whitedwarfs2}.  The progenitors of these
white dwarfs presumably formed as a very early generation of stars. This
scenario has its own problems, however.  The progenitor stars would have
produced copious amounts of metal enriched gas, which is not seen in the
Galactic abundances \cite{metals,metals2}.  Ridding the Galaxy of this
metal enriched gas would require a strong Galactic wind at the appropriate
time \cite{brianmodel}.  In addition, this scenario requires a stellar
mass function for the halo stars that is peaked at a much higher mass than
is observed in the disk today.  Unless the initial mass function is
strongly suppressed for low masses we should still see the low mass main
sequence stars that would have been produced along with the higher mass
white dwarf progenitor stars \cite{IMFproblem,IMFproblem2}.  Primordial
black hole candidates require a fine tuning of the initial distribution of
density perturbations, although there has been some recent work on black
holes formed at the QCD phase transition \cite{Jedamzik}.

In this paper we explore an alternative explanation for the relatively
long term event durations seen by the MACHO group.  MACHO mass estimates
are derived from the observed event duration, with the assumption of a
model for the MACHO distribution and velocity structure.  The event
duration is a function not only of the mass of the lens, but also of the
distance to the lens and its tranverse (with respect to the line of sight)
velocity.  Without the additional information obtained, for example, by a
parallax or binary lens event, in which case the velocity of the lensing
object can sometimes be extracted and an independent estimate of the
distance to the lens, the mass of an individual event cannot be determined
directly. The distance along the lensing tube and tranverse velocity are
known only in a statistical sense (for a given model), and therefore only
a statistical estimate of the mass of a population of lenses can be made.
However, even given a large number of events, the mass estimate still
depends strongly on the assumed distribution function (phase-space
density) of the lenses.  For example, MACHOs which are moving through the
lensing tube with a transverse velocity that is much less than that in a
standard isothermal halo model will result in much longer event durations
for a given mass.

Brown dwarf stars, with masses below $0.08 M_{\odot}$, represent a less
troublesome candidate for the observed microlensing events toward the
Large Magellenic Cloud (LMC) but appear incompatible with the current data
in the context of a standard isothermal halo model \cite{MACHOmass}.
However, the viability of such lens candidates could be very different in
a model with a non-standard MACHO distribution function.

In order to explore the effects of a non-standard MACHO distribution
function, we allow for a bulk rotational component of the MACHO halo and
anisotropic velocity dispersion.  While a large bulk rotation of a
spherical halo seems unlikely, there is no reason to exclude such a
component {\it a priori}.  Further, the MACHOs, which represent only about
$20\% - 40\%$ of the total halo mass \cite{ggt,ggt2}, may be in a more
condensed or flattened distribution than the CDM halo, and thus it seems
reasonable to expect some rotational component to their velocity profile.

\section{Microlensing Tools}
	
In order to compare our models to the MACHO data we first calculate the
average duration. Note that we are not attempting here to determine a
detailed model for the halo, in which case we would need to consider the
event duration on an event-by-event basis, but rather to explore the
potential of non-standard halo distribution functions to affect the MACHO
mass estimate.  The average event duration is directly related to the
optical depth, rate and MACHO mass for a given model.
\be
\bar{t}_{model}=\frac{\tau}{\Gamma},
\ee
or 
\be
\bar{t}_{model}=\frac{\tau_{model}}{\Gamma_{1M_\odot}}
\sqrt{\frac{\overline{m}}{1 M_\odot}}
\ee
where we have now calculated the rate for solar mass MACHO's and made the
mass dependence explicit. 
For a given model we
can then extract the prediction for the average MACHO mass,
\be
\frac{\overline{m}}{1 M_\odot}= \left[\bar{t}_{obs}
\frac{\Gamma_{1M_\odot}}{\tau_{model}}\right]^2.
\ee

Calculation of the observed average duration (Appendix A) gives a value of
61 days. Due to the low number of events and the statistics of
microlensing \cite{HanGould}, the possible error is very
large. Nevertheless, for the remainder of this paper we will utilize the
observed average duration: we are interested in how rotation of the halo
might effect the {\em central} value for the MACHO mass estimate.

The optical depth towards the LMC is evaluated in the standard way,
\be
\tau = \pi \int_0^{D_s}  R_E^2(l) \frac{\rho(l)}{m} dl
\ee
where $R_E(l)$ is the Einstein radius at distance $l$, 
\be
R_E(l)= \sqrt{\frac{4 G m}{c^2} \frac{l(D_s-l)}{D_s}},
\ee
and $D_s=50\,\kpc$ is the distance to the source.

Calculating the microlensing rate for a moving observer and source (the
microlensing tube is tumbling) in an arbitrary MACHO distribution is
slightly more complex. For each segment $dl$ of the microlensing tube we
can calculate the corresponding infinitesimal contribution to the rate,
$d\Gamma$. Let $f_T(v,l)$ be the distribution function of the MACHOs in
the frame of the tube at location $l$ along the tube. For every element of
the velocity distribution $\Delta v_x \Delta v_y \Delta v_z$ there is a
contribution to the rate,
\begin{eqnarray}
\Delta^3 d\Gamma &=& 2 R_E(l) dl \abs{\vec{v}_\perp} f_T(\vec{v},l) \Delta
v_x \Delta v_y \Delta v_z \nonumber \\
                 &=& 2 R_E(l) dl \abs{\vec{v} \times \hat{l}}
f_T(\vec{v},l) \Delta v_x \Delta v_y \Delta v_z
\end{eqnarray}
where $2 R_E(l) dl$ is the cross-section of the segment and $\hat{l}$ is
the unit vector pointing along the microlensing tube. Integrating over all
velocities we have
\begin{eqnarray}
d\Gamma &=& 2 R_E(l) dl \int_{\vec{v}}^{} \abs{\vec{v} \times \hat{l}}
f_T(\vec{v},l) d^3\vec{v}. \nonumber \\
    &=& 2 R_E(l) dl <\abs{v_{\perp}}> \rho(l)/m,
\end{eqnarray}
where $<\abs{v_{\perp}}>$ is the expectation value of the transverse velocity
relative to the microlensing tube. Integrating along the line of sight we
obtain
\be
\Gamma = 2 \int_0^{D_s} R_E(l) \frac{\rho(l)}{m}<\abs{v_{\perp}}> dl.
\ee

It will be useful to consider the average duration due to events at a
distance $l$ from the Sun,
\begin{eqnarray}
\bar{t}_{0.1}(l) &=& \frac{d\tau(l)}{d\Gamma(l)} \nonumber \\
		 &=& \frac{\pi R_E(l)}{2 <\abs{v_\perp}>},
\end{eqnarray}
where the Einstein radius is evaluated assuming a mass of $0.1\Msol$.  We
also consider the fractional event rate, $d\Gamma/\Gamma$, as a function
of the distance $l$, i.e. the fraction of events that occur at any given
distance.

\section{Effect of Rotation}

We start by examining the ``standard'' halo model,
\be
\rho = \rho_0 \frac{r_0^2 + a ^2}{r^2+a^2} \\
\ee
where $a=5\,\kpc$ and $r_0=8.0\,\kpc$, with no rotation, flattening or
anisotropy. Figure 1 shows the transverse velocity with respect to the
microlensing tube as a function of distance from the Sun. The transverse
velocity is largest towards the ends where the tumbling velocity due to
the motion of the sun and LMC is greatest, but overall is essentially
flat. Figure 2 shows the average event duration, again as a function of
distance, assuming a MACHO halo composed of $0.1\Msol$ objects. The
duration is largest near the center and drops to zero for objects close to
either end of the microlensing tube. The maximum average duration is quite
low, only about 43 days.  Regardless of the distribution of event
distances, the average duration over the entire tube can be no larger than
this. Scaling to the observed average duration of 61 days we find a
minimum possible average mass of $0.2\Msol$.  The fractional event rate as
a function of distance for this model is shown in Figure 3.  The actual
distribution of events favours the first part of the microlensing tube,
where the Einstein radius is small and durations correspondingly short,
leading to a much larger estimated MACHO mass, $\sim 0.4\Msol$.  Since the
transverse velocity is relatively flat over the entire distance to the
LMC, this distribution is driven mainly by the distribution of the lenses.

\begin{figure}[htb]
\epsfysize=10.0cm
\centerline{
\rotate[r]{\epsfbox{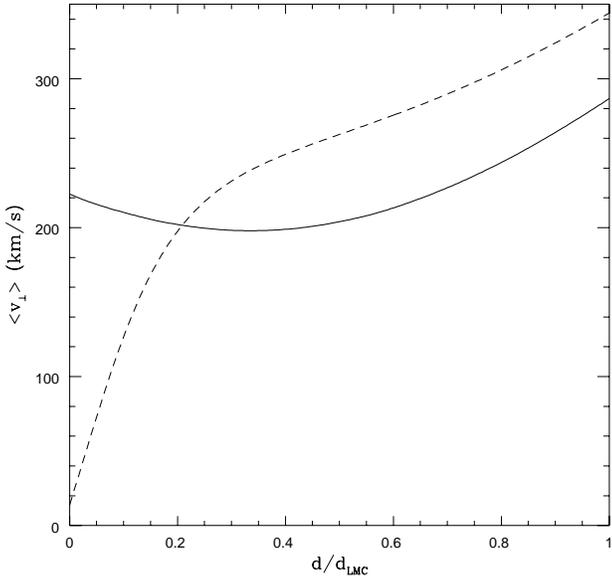}}}
\caption{Transverse velocities for the ``standard'' (solid) and rotating
(dashed) halos.}
\end{figure}

\begin{figure}[htb]
\epsfysize=10.0cm
\centerline{
\rotate[r]{\epsfbox{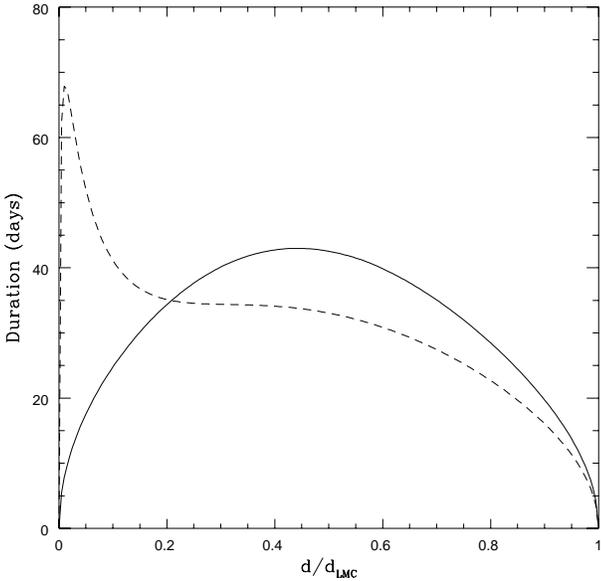}}}
\caption{Average event durations assuming $0.1\Msol$ MACHOs for the 
``standard'' (solid) and rotating
(dashed) halos.}
\end{figure}

\begin{figure}[htb]
\epsfysize=10.0cm
\centerline{
\rotate[r]{\epsfbox{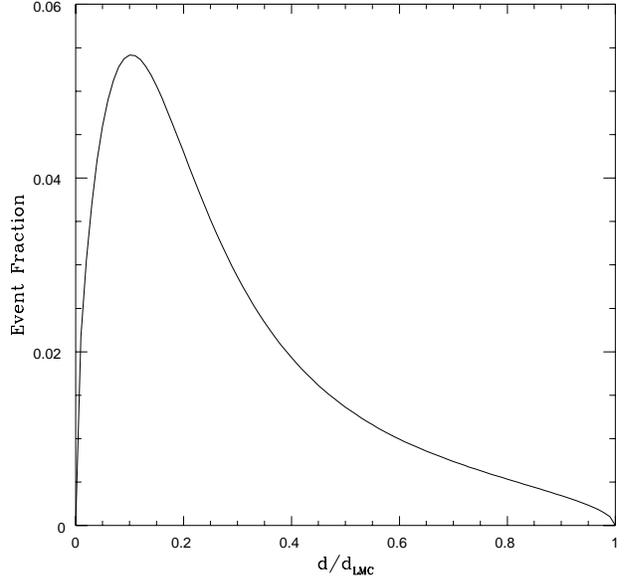}}}
\caption{Distribution of the event distances for the ``standard'' halo.}
\end{figure}

Next, for comparison, we consider an extreme model -- an idealized cold
co-rotating halo ($v_{\rm rot}$=220 km/s) model with zero velocity
dispersion (even in the z-direction). While obviously not a realistic
model, it does allow us to examine the limits on MACHO mass estimate
reduction in models with a rotational component for the halo velocity
structure.  Figure 1 shows the average transverse velocity of the this
model as a dashed line. As expected, the transverse velocity close to the
observer is low. However, note that the transverse velocity at more
distant points is actually quite high: the MACHOs sweep past the
microlensing tube because the LMC velocity is not aligned with the
rotation of the Milky Way as is the motion of the Sun.  The dashed line in
Figure 2 is the average duration of events (for $0.1\Msol$ MACHOs) for the
co-rotating halo model. We see a dramatic effect due to the reduced
velocities close to the Sun: there is a sharp peak in the durations of
events at very small distances. In general, the durations are increased
for distances smaller than 10 kpc and decreased for distances greater than
10 kpc, reflecting the relative velocities at these distances.  There is a
small region where the durations are above 61 days, the observed average
duration. This implies that it is possible, at least in principle, to
obtain average durations matching those observed with $0.1\Msol$ MACHOs,
and suggests distributions which are highly concentrated toward the center
of the Galaxy as the most likely way to achieve this.

If a rotating halo is to lead to reduced mass prediction for the MACHOs,
essentially all the lensing must occur very close to the observer.  The
distribution from the standard model is too wide and peaks at too large a
value to fully sample the initial peak in average event duration due to
rotation. Any benefit from the increased durations at short distance will
be offset by the decreased durations at larger distances.  We consider two
options for bringing the microlensing closer to home: a density fall-off
faster than $1/r^2$ or a flattened halo. Both of these posibilities are
quite reasonable. The stellar halo population we {\em do} know about, the
spheroid, has a density distribution that falls off roughly as
$1/r^{3.5}$. The MACHO halo may trace this distribution. On the other
hand, various observations \cite{Olling,Sackett} suggest that dark halos
are flattened, and simulations consistently show that the CDM halos
generically form with departures from spherical symmetry. The dissipative
baryonic component of the halo, the MACHOs, might be expected to be even
further from spherical.  We know of at least one component of the Galaxy,
the disk, that became significantly flattened before forming stars. In the
following section we examine these models in detail.

\section{Modeling}

Following the above discussion, we will consider three density profiles: 

\noindent an isothermal halo model with core radius $a$
\be
\rho = \rho_0 \frac{r_0^2 + a ^2}{r^2+a^2},
\ee
a flattened (axis ratio q=0.2) inverse square halo \footnote{
We considered using an Evans model for the halo since these come with a
consistent prescription for the phase space density. However such
models cannot produce a halo as flattened as we wished to explore, at most
obtaining an equivalent flattening in the density of $\approx 1/3$.}

\be
 \rho =  \rho_0 \frac{r_0^2 + a ^2}{R^2 + z^2/q^2 +a^2}, \label{flathalo}
\ee
and a spheroid-like halo
\be
\rho = \rho_0 \frac{r_0^{3.5}+a^{3.5}}{r^{3.5}+a^{3.5}} .
\ee

We fix the core radii at $a=5$ kpc for the $1/r^2$ models and $a=1 $kpc
for the spheroid model.  (We examined the effects of varying the core
radius for all 3 models and found them to be unimportant to our final
results.) 

To estimate the MACHO mass implied by these models, we need to specify not
only the density distribution, but also the velocity structure of the
halo. We assume that the phase space distribution is a simple anisotropic
gaussian, offset from the origin by the rotation velocity and aligned at
each point with cylindrical Galactic coordinates.\footnote{Most of the
lensing in our models takes place close to the Solar position, where a
velocity ellipsoid aligned with Galactic cylindrical coordinates is a
reasonable approximation, and over a relatively narrow range in distance
where the anisotropic velocity dispersions should not change radically.
Thus, while these are not the most general set of assumptions, they are
sufficient for our purposes and greatly simplify the calculations.} We do
not attempt to construct completely ``self-consistent'' halo models. For
the flattened or centrally condensed halos we consider, such models would
need to incorporate the bulge and disk, as well as halo
self-consistency. This is presently well beyond the state of the art given
the large uncertainties in essentially all Galactic
parameters. Furthermore, microlensing constraints are sufficiently weak
(due to low number of events) that such sophistication is unwarranted at
present.

The Jeans equations for the above time-independent distributions simplify to,
\begin{eqnarray}
\ptlp{\rho\overline{v_R^2}}{R} +
\frac{\rho}{R}(\overline{v_R^2}-\overline{v_\phi^2}) + \rho \ptl{\Phi}{R} =0
\\
\ptlp{\rho\overline{v_z^2}}{z} + \rho\ptl{\Phi}{z} = 0,
\end{eqnarray}
where $\Phi$ is the potential of the total halo.  We further assume that
$\ptl{\overline{v_R^2}}{R}\approx 0$ and $\ptl{\overline{v_z^2}}{z}\approx
0$ at points in the halo of interest for this work.  Although this is
probably a poor assumption for detailed modeling, we accept it in order to
focus on the effects of {\em rotation} on the microlensing mass
estimates. We then have
\begin{eqnarray}
\left( -\ptl{\ln\rho}{\ln R} - 1\right)\sigma_R^2 +
\sigma_\phi^2+\overline{v_\phi}^2 &=& R \ptl{\Phi}{R} \nonumber \\
\left(\ptl{\rho}{z}\right) \sigma_z^2 &=& - \rho \ptl{\Phi}{z} 
\end{eqnarray}
where $\sigma_R^2=\overline{v_R^2}$, $\sigma_z^2=\overline{v_z^2}$ and
$\sigma_\phi^2 + \overline{v_\phi}^2=\overline{v_\phi^2}$ (breaking it up
into random and bulk flows).  The derivative $\ptl{\Phi}{R}$ is just the
R-component of the force at R.  Since we are roughly at the equator,
$R\ptl{\Phi}{R}$ is simply the square of the circular velocity at R,
$v_c^2(R) \approx (220 {\rm km/s})^2$

To complete this set of equations we need to find $\ptl{\Phi}{z}$. We
assume that the bulk of the halo, and thus the major contribution to the
potential, comes from the non-baryonic component, whose distribution is
assumed to be that of equation \ref{flathalo}, with core radius $a_{\rm
NB} = 0$ and arbitrary flattening $q_{\rm NB}$ independent of the MACHO
halo. It can be shown (Appendix B) that to first order in $z$ 
\be
\ptl{\Phi}{z}= v_c^2 \left[\sqrt{\frac{q_{\rm NB}^2}{1-q_{\rm
NB}^2}}sin^{-1}\sqrt{1-q_{\rm NB}^2}\right]^{-1} \frac{z}{R^2}
\ee
near the equatorial plane.  This gives
\be
\sigma_z^2 = - \rho  v_c^2 \left[\sqrt{\frac{q_{\rm NB}^2}{1-q_{\rm NB}^2}}
sin^{-1}\sqrt{1-q_{\rm NB}^2}\right]^{-1}
\frac{z}{R^2} \left(\ptl{\rho}{z}\right)^{-1}.
\ee
The lowest value for $\sigma_z^2$ is obtained for a spherical (q=1.0)
non-baryonic halo. A larger value of the vertical velocity dispersion will
drive the transverse velocities up, the event durations down, and
ultimately yield a higher mass estimate. Since we are interested here in
models which lower the MACHO mass estimates we take $q_{\rm NB} = 1$ and
find
\begin{eqnarray}
\left( -\ptl{\ln\rho}{\ln R} - 1\right)\sigma_R^2 +
\sigma_\phi^2+\overline{v_\phi}^2 &=& v_c^2
\label{vsigrelation} 
\\
\nonumber
\left(\ptl{\rho}{z}\right) \sigma_z^2 &=& -\rho v_c^2 \frac{z}{R^2}
\end{eqnarray}

As we will see, the precise form of the relationship between the
rotation and velocity dispersion is not important for our conclusions. 
We use equation \ref{vsigrelation} as a general
guide to which combinations of rotation and velocity dispersions are
reasonable. 

\section{Results and Discussion}
We plot our results for the various halo models in figures 4 through 6.
The solid contours correspond to the predicted MACHO mass estimate,
corresponding to the average MACHO event duration of 61 days, as a
function of the (one-dimensional) velocity dispersion \footnote{In the
discussion that follows we take $\sigma= \sigma_R = \sigma_\phi$. We have
also considered models with a constant anisotropy $\sigma_R = \alpha
\sigma_\phi$, where $0.3\le\alpha\le 3.0$; our results are essentially
unchanged for these models.}  and rotation speed.  For reference, the
standard non-rotating halo with isotropic maxwellian velocity dispersion
has a one-dimensional dispersion $\sigma =156 {\rm km/s}$.

The models within $\pm 10\%$ of our $\sigma-v_{\rm rot}$ curve (equation
\ref{vsigrelation}) lie in the shaded region between the dotted lines.
Models below this region are unlikely to have sufficient support. The
trade-off of dispersion velocity for rotation velocity can be seen
clearly.

\begin{figure}[htb]
\epsfysize=10.0cm
\centerline{
\rotate[r]{\epsfbox{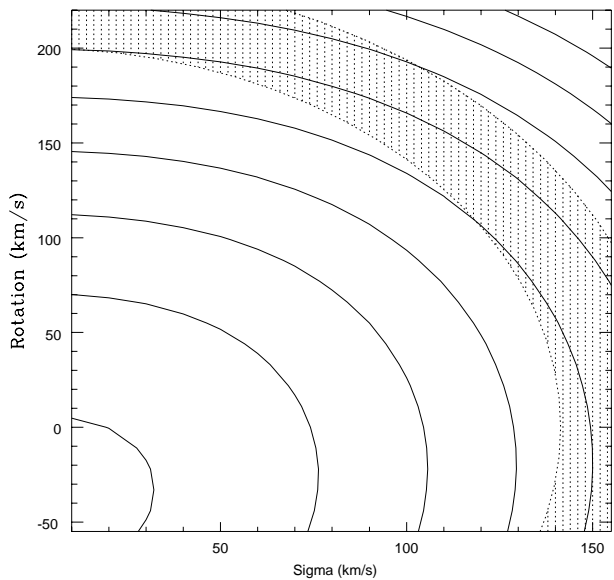}}}
\caption{Average MACHO mass for an isothermal halo based on the MACHO
collaboration average event duration. Countours are 0.15, 0.2,
0.25, 0.3,..$\Msol$ from left side.}
\end{figure}

Figures 4, 5 and 6 show the predicted MACHO masses for the spherical
$1/r^2$, flattened $1/r^2$ and spherical $1/r^{3.5}$ halo models.  In no
case does the predicted mass for the current MACHO event duration go below
$0.25\Msol$ in the allowed region indicated by equation \ref{vsigrelation}
. As expected following the discussion in section 3, the spherical $1/r^2$
model actually predicts increasing MACHO mass estimates as it becomes more
rotation supported. Rotation increases the velocities along the
microlensing tube far from the Sun leading to shorter event durations for
a given MACHO mass, or conversely, a larger MACHO mass estimate for a
given observed event duration.  Most microlensing in the standard halo
model occurs at distances where the velocities are increased when a
rotational velocity component is added and hence the mass estimates for a
given event duration increase.

\begin{figure}[htb]
\epsfysize=10.0cm
\centerline{
\rotate[r]{\epsfbox{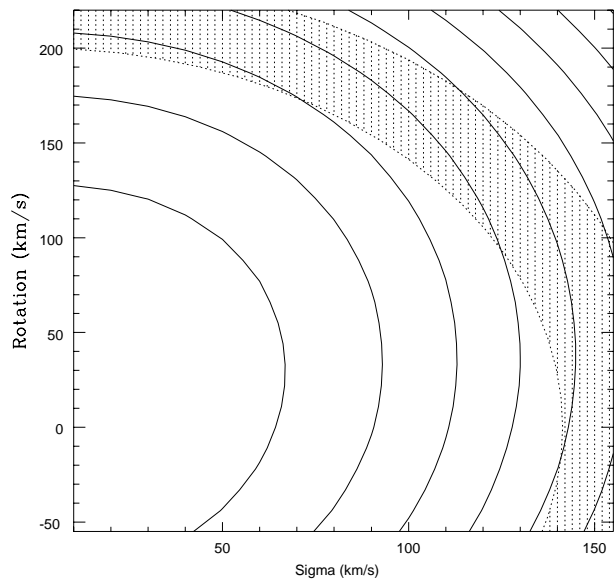}}}
\caption{Same as Figure 4 for the flattened $1/r^2$ halo. Countours are
0.15, 0.2,
0.25, 0.3,..$\Msol$ from left side.}
\end{figure}

In contrast, in highly flattened $1/r^2$ halo models rotation does lead to
reduced mass estimates.  However, the improvement is not as dramatic as
might be hoped. The reason is a little subtle. In the absence of rotation,
the microlensing is concentrated in the first part of the tube by the
flattening of the halo. We can see that predicted mass estimates for the
flattened halo model with no rotation are a little higher than for the
spherical model since in this region the durations for the non-rotating
model are short.  However, the fractional rate is also dependent on the
average transverse velocity of MACHOs. Because a flattened halo has a
lower z-velocity dispersion, when rotation is added the transverse
velocity can become very low close to the observer. This is precisely why
rotation might be expected to lower the predicted MACHO mass
estimates. However, the low transverse velocity also suppresses the event
rate close to the observer and shifts the distribution of events to larger
distances where the transverse velocities are higher. These two effects
compete, with the end result that the reduction to the MACHO mass estimate
due to rotation is only modest.

\begin{figure}[htb]
\epsfysize=10.0cm
\centerline{
\rotate[r]{\epsfbox{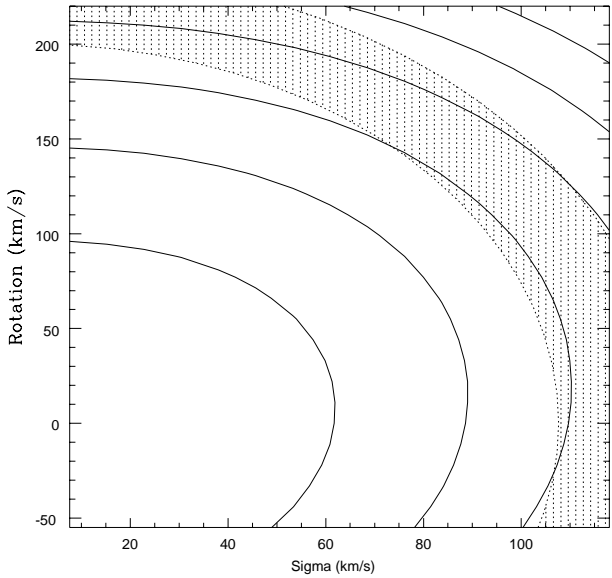}}}
\caption{Same as Figure 4 for the spheroidal halo. Countours are 0.2,
0.25, 0.3,..$\Msol$ from left side.}
\end{figure}

Our second hope for taking advantage of rotation, a more concentrated
$1/r^{3.5}$ spheroid, is also disappointing. Supporting a spheroidal halo
rotationally leads to very little change in the MACHO mass estimate.
Rotation fails to reduce the mass estimates in the spheroidal case for
reasons slightly different than those for the flattened halo. For a
spheroidal MACHO distribution, transverse velocities never get very low,
even close to the observer.  Hence, little benefit is derived from the
central concentration.  Such a spheroidal model also has serious
difficulties producing a high optical depth without ruining the flatness
of the Galactic rotation curve.

In this section we have explored three general classes of halo models,
which have been unable, even with the additional of a generous rotational
component, to reduce the MACHO mass estimates for the current observed
event durations to masses in the substellar regime. The failure of
rotating halos to significantly reduce MACHO mass estimates can be summed
up as follows.  First, the virial theorem implies that a typical velocity
for a particle in an extended halo will be roughly the same regardless of
whether the halo is or is not rotationally supported.  Velocities cannot
be made arbitrarily low.  Secondly, the direction to the LMC is
``awkward'', that is, it respects none of the symmetries of the Galaxy.
This has several implications for attempts to lower the MACHO mass
estimates.  Bulk velocities due to rotation cannot be used to cancel out
some of the motion of the MACHOS through the lensing tube except possibly
very close to the observer.  An effort to concentrate the MACHOs in this
region, however, is self-defeating in the models we explored.  The event
rate in such models is dominated by MACHOs moving through the tube further
away from the observer, and the net effect on the MACHO mass estimate is
not sufficiently large to reach the substellar regime with current data.
Further, the direction of the LMC makes it impossible to arrange, in any
natural way, an anisotropic velocity dispersion such that the net
transverse velocity through the tube is significantly reduced
\cite{anisotropy}.

\section{Conclusions}

Scenarios in which brown dwarfs populate a standard spherical,
non-rotating halo in significant numbers predict event durations
significantly (more than a factor of 2) shorter than observed in the
current MACHO data. We have explored halo models with a bulk rotational
component in an attempt to lower the predicted mass estimate for the MACHO
events to the substellar regime.  We find that unless essentially {\em
all} of the lensing takes place within 1-2 kpc of the Sun {\em and}
residual velocity dispersions are very low, $\lta 30 {\rm km/s}$, it is
not possible to reduce the MACHO mass estimate in these models to the
brown dwarf mass range.  Such a configuration no longer resembles a halo
distribution for the MACHOs.

A highly flattened rotating halo was our most successful model in reducing
MACHO mass estimates, although even this model did not reduce the
microlensing mass estimate below about $0.25\Msol$.  This suggests that a
thick disk configuration with a steeper (exponential) density fall off
away from the Galactic plane might be able to reduce the mass estimates
further. A high rotation velocity is also more natural for such a
distribution than for a more extended halo. Preliminary work shows that a
thick disk can indeed reduce the mass estimate, but only to around
$0.15-0.2\Msol$. We will report on this in more detail in a forthcoming
paper.

In summary, the prospect for a brown dwarf microlensing halo
appears dim given the current data, and the question of what and
where the MACHO lenses are remains unanswered.

\section*{Appendix A : Calculation of Average Event Duration}
{}From the observed events we can construct the {\em
observed} distribution function of time scales, $\nu(\hat{t})$. The true
distribution will then be $\rho(\hat{t})=\nu(\hat{t})/\epsilon(\hat{t})/\int
\nu(\hat{t})/\epsilon(\hat{t}) d\hat{t}$. We want to calculate
\begin{eqnarray}
\bar{\hat{t}} & = & \int \hat{t} \rho(\hat{t}) d\hat{t}  \\
        & = & \frac{\int \hat{t} \nu(\hat{t})/\epsilon(\hat{t}) dt}
{\int\nu(\hat{t})/\epsilon(\hat{t}) d\hat{t}}
\end{eqnarray}
which requires knowing the function $\nu(\hat{t})$. Our best guess for
this will be to let
\be
\nu(\hat{t})=\frac{1}{n}\sum_i \delta(\hat{t}_i - \hat{t}).
\ee
Substituting in
gives 
\be
\bar{\hat{t}}= \frac{\sum_i \frac{\hat{t}_i}{\epsilon(\hat{t}_i)}}
{\sum_i \frac{1}{\epsilon(\hat{t}_i)}}.
\ee
Using the events and efficiencies towards the LMC \cite{MACHOmass} we
obtain $\bar{\hat{t}}= 61$ days.


\section*{Appendix B : Vertical Potential}

Following \cite{binney}, the $z$-component of the
gravitational force 
generated by a density distribution
$\rho=\rho(m^2)$ where $m^2=R^2+z^2/q^2$, is given by
\be
-\ptl{\Phi}{z} = 4 \pi G \sqrt{1-e^2} \int_0^R \frac{\rho(m^2) z dm}{m
[{R^2\over m^2} - (1-q^2)]^{3/2}},
\ee
where we keep only the leading order term in z, in order to obtain a lower
limit on the estimate for $\sigma_z$.

Assuming a density $\rho=\rho_0 m_0^2/m^2$ we then arrive at
\be
-\ptl{\Phi}{z} = 4 \pi G \frac{\rho_0 m_0^2 z}{R^2}
\ee
Rewriting this in terms of the circular velocity and flattening we find
\be
\ptl{\Phi}{z} = \frac{v_c^2}{\sqrt{q^2/(1-q^2)} \sin^{-1}(1-q^2)}\frac{z}{R^2}.
\ee

\end{document}